\documentclass[12pt,a4paper]{article}
\usepackage[latin1]{inputenc}
\usepackage{amsmath}
\usepackage{amsfonts}
\usepackage{amssymb}
\usepackage{graphicx}
\usepackage{epsf,color}
\usepackage{authblk}

\begin{document}

\author[1,2]{William N. Plick}
\author[1,2]{Radek Lapkiewicz}
\author[1,2]{Sven Ramelow}
\author[1,2,3]{Anton Zeilinger}
\affil[1]{\small\emph{Quantum Optics, Quantum Nanophysics, Quantum Information, University of Vienna, Boltzmanngasse 5, Vienna A-1090, Austria}}
\affil[2]{\emph{Institute for Quantum Optics and Quantum Information, Boltzmanngasse 3, Vienna A-1090, Austria}}
\affil[3]{\emph{Vienna Center for Quantum Science and Technology, Faculty of Physics, University of Vienna, Boltzmanngasse 5, Vienna A-1090, Austria}\normalsize}

\title{The Forgotten Quantum Number: A short note on the radial modes of Laguerre-Gauss beams}

\maketitle

\begin{abstract}
\noindent The orbital angular momentum quantum number of Laguerre-Gauss beams has received an explosively increasing amount of attention over the past twenty years. However, often overlooked is the so-called radial number of these beams. We present a derivation of the differential operator formalism of this ``forgotten" quantum number. We then draw some connections between this new formalism and the effect the radial number has on beam stability with possible application to quantum communication. We also briefly outline how the radial number is tied to other physical aspects of the beam (such as the Gouy phase, and radial confinement). These do not necessarily constitute finished results, but are instead meant to stimulate discussion of this interesting and often overlooked physical parameter.
\end{abstract}

\section{Introduction}

\noindent In 1992 Allen et. all \cite{A} elucidated the connection between the azimuthal mode number \emph{l} of Laguerre-Gauss (LG) beams and the concept of the orbital angular momentum (OAM) of light. Since then this parameter has been the subject of a flurry of attention \cite{rev}. This activity has been fueled by interest in several technological applications, including micro-mechanical manipulation, atom-trapping, and not the least, using the OAM quantum number as a high dimensional Hilbert space for tasks in quantum and classical information transfer. Furthermore such inquires have helped to expand our understanding of the fundamental nature of light (see Ref.\cite{rev} and references therein).\\

\noindent Despite all this activity, very little attention has been payed to the other mode number of LG modes \--- the so-called ``radial number". This moniker is due to the fact that LG beam posses $n$, or $n+1$, rings in the transverse amplitude profile \--- depending on whether the beam's OAM is zero, or non-zero, respectively. However, the deeper physical meaning behind this value has not been clear. We seek a better understanding of this quantity.\\

\noindent To begin we write the equation of a LG beam in circular-cylindrical coordinates ($r$, $\phi$, $z$) for reference,

\begin{eqnarray}
\mathrm{LG}_{nl}(r,\phi ,z)&=&\sqrt{\frac{4n!}{\pi (n+l)!}}\frac{1}{w_{z}}\left(\frac{\sqrt{2}r}{w_{z}}\right)^{l}L_{n}^{l}\left(\frac{2r^{2}}{w_{z}^{2}}\right)\mathrm{exp}\left[-\frac{r^{2}}{w_{z}^{2}}\right]\\
& &\times\mathrm{exp}\left[i\left(\frac{kr^{2}}{2R_{z}}-(2n+l+1)\varphi_{g}\right)\right].\label{LG}
\end{eqnarray}

\noindent Where $n$ and $l$ are the radial and orbital angular momentum quantum numbers, respectively; $L_{n}^{l}$ is the generalized Laguerre polynomial of order $n$ and degree $l$. The functions $w_{z}$, $R_{z}$, and $\varphi_{g}$ are the beam waist, radius of curvature, and Gouy phase of the fundamental beam, and are given by

\begin{eqnarray}
w_{z}&=&w_{o}\sqrt{1+\frac{4z^{2}}{k^{2}w_{o}^{4}}},\\
R_{z}&=&z+\frac{k^{2}w_{o}^{4}}{4z},\\
\varphi_{g}&=&\mathrm{arctan}\left[\frac{2z}{kw_{o}^{2}}\right].\label{gouy}
\end{eqnarray}

\noindent Where $k$ is the overall wavenumber of the beam, and $w_{o}$ is the beam waist at $z=0$ (defined where the beam is narrowest). The beam or photon is then completely defined by four numbers $n$, $l$, $k$, $w_{o}$. LG beams are solutions to the paraxial wave equation (PWE)

\begin{eqnarray}
\nabla^{2}_{t}E-2ik\frac{\partial}{\partial z}E=0.\label{PWE}
\end{eqnarray}

\noindent Where $\nabla^{2}_{t}$ is the transverse Laplacian and $E$ is a complex electric scalar field (i.e. we assume polarization is uniform).\\

\noindent The definition of the differential OAM operator, about the direction of beam-propagation, is well known and straightforward

\begin{eqnarray}
\hat{L}_{z}=-i\frac{\partial}{\partial\phi}.
\end{eqnarray}

\noindent Where, for now, we have set $\hbar=1$. \\

\noindent In the following section we develop the differential operator formalism for the \emph{other} mode number which governs the spatial profile of the LG photons, the radial quantum number ($n$-modes). In section three we draw some connections between the radial quantum number and other physical quantities, including some speculation about how the $n$-modes may be utilized for robust quantum communication protocols. In section four we summarize and conclude.

\section{The differential operator formalism of the radial quantum number}

\noindent In order to derive the differential operator formalism of the $n$-modes we start with the namesake of the LG beams: the Laguerre polynomial. There exist a series of relations between Laguerre polynomials of varying order and degree. One such relation is

\begin{eqnarray}
nL_{n}^{l}(x)=(l+1-x)L_{n-1}^{l+1}(x)-xL_{n-2}^{l+2}(x).
\end{eqnarray}

\noindent Which, when combined with the rule for differentiation of the polynomials,

\begin{eqnarray}
\frac{\partial}{\partial x}L_{n}^{l}(x)=-L_{n-1}^{l+1}(x),
\end{eqnarray}

\noindent yields,

\begin{eqnarray}
\left[(x-l-1)\frac{\partial}{\partial x}-x\frac{\partial^{2}}{\partial x^{2}}\right]L_{n}^{l}(x)=nL_{n}^{l}(x).\label{L}
\end{eqnarray}

\noindent Given this differential relation, it is possible to arrive at a relationship between the full LG modes by left-multiplying the other factors (the non-polynomial terms) in the LG function Eq.(\ref{LG}) onto Eq.(\ref{L}), and commuting those factors past the differentials on the left. Doing this we obtain

\begin{eqnarray}
\left[-\frac{1}{4x}\frac{\partial^{2}}{\partial\phi^{2}}-x\frac{\partial^{2}}{\partial x^{2}}+\frac{i}{2}\frac{\partial}{\partial\phi}-\frac{\partial}{\partial x}+\frac{x}{4}-\frac{1}{2}\right]\mathrm{LG}_{nl}(x,\phi)=n\mathrm{LG}_{nl}(x,\phi).
\end{eqnarray}

\noindent Now if we make the the identification $x=2r^{2}/w_{o}$ and perform a change of coordinates, we obtain

\begin{eqnarray}
\hat{N}_{o}\equiv\left[-\frac{w_{o}^{2}}{8}\left(\frac{1}{r^{2}}\frac{\partial^{2}}{\partial\phi^{2}}+\frac{1}{r}\frac{\partial}{\partial r}+\frac{\partial^{2}}{\partial r^{2}}\right)+\frac{i}{2}\frac{\partial}{\partial\phi}+\frac{1}{2}\left(\frac{r^{2}}{w_{o}^{2}}-1\right)\right].\label{comp}
\end{eqnarray}

\noindent Where, $\hat{N}_{o}L_{n}^{l}(r,\phi,0)=nL_{n}^{l}(r,\phi,0)$, and we have defined the differential $n$-mode operator for $z=0$. An entirely equivalent operator to Eq.(\ref{comp}) has been derived previously by Karimi and Santamato \cite{karimi}. They arrived at their result be examining the group algebra of the LG modes. It is compelling that the two dissimilar derivations yield the same result.\\

\noindent Further simplification is possible by identification with other, better known, operators yielding the opportunity for some physical insight

\begin{eqnarray}
\hat{N}_{o}=-\frac{w_{o}^{2}}{8}\nabla^{2}_{t}-\frac{\hat{L}_{z}}{2}+\frac{1}{2}\left(\frac{r^{2}}{w_{o}^{2}}-1\right).\label{N0}
\end{eqnarray}

\noindent The operator is composed of the transverse laplacian and thus has a direct relationship to the transverse momentum (up to some constants). It also contains the OAM operator (again, with $\hbar=1$), as well as a term reminiscent of a harmonic potential. It commutes with the OAM operator, but not with the Hamiltonian for free propagation due to the final potential-like term. This indicates that $n$ is not a conserved quantity of the beam with respect to evolution through free space. This point will be discussed further in the following sections. The operator is also Hermitian.\\

\noindent It is also useful to derive a general operator for any value of $z$. Continuing with the same methods as above we have

\begin{eqnarray}
\hat{N}_{z}=-\frac{w_{z}^{2}}{8}\nabla^{2}_{t}+\frac{iz}{2}\frac{\partial}{\partial r}r-\frac{\hat{L}_{z}}{2}+\frac{1}{2}\left(\frac{r^{2}}{w_{o}^{2}}-1\right).
\end{eqnarray}

\noindent The differences being that the pre-factor of Laplacian is now the $z$-dependent beam waist and an additional term resultant from the phase imparted from the radius of curvature, $R_{z}$ (this can be seen from the fact that if the LG beam is written without the radius of curvature phase factor then this term does not appear). Interestingly, the final term is unchanged and remains $z$-independent.

\section{Physical implications and connections}

\noindent Now that we have a well defined operator that has the LG modes as its eigenstates with the integer values from 0 to $\infty$ as its eigenvalues, we can begin to ask what physical implications this formalism may have. To start, if we are to believe that, $\hat{N}_{z}$ is a proper quantum mechanical observable (as is suggested by it being Hermitian), then it should follow that it has a conjugate variable. Since it is not entirely obvious what this conjugate is from symmetry considerations (an example of an obvious conjugate pair would be the OAM and angular position), we can naively attempt to find the conjugate by ``brute force" in the following way. We define a function of the $r$ and $z$ position operators as $f(r,z)$. Then we demand that $\left[\hat{N}_{z},f(r,z)\right] g(r,z)=a*g(r,z)$, where $g(r,z)$ is an arbitrary test function, and $a$ is an arbitrary constant (from now on $a=1$ for simplicity). Making a substitution of the Laplacian using the PWE (\ref{PWE}), this can be re-written as a differential equation which after some simplification is

\begin{eqnarray}
\left(\frac{kw_{o}^{2}}{4}+\frac{z^{2}}{kw_{o}^{2}}\right)\frac{\partial f}{\partial z}+\frac{zr}{kw_{o}^{2}}\frac{\partial f}{\partial r}=1.
\end{eqnarray}

\noindent Where the function $g$ has canceled out. The general solution of this equation is

\begin{eqnarray}
f(r,z)=-2\mathrm{arctan}\left[\frac{2|z|}{kw_{o}^{2}}\right]+C_{1}\mathrm{ln}\left[\frac{k^{2}w_{o}^{4}+4z^{2}}{4r^{2}}\right].
\end{eqnarray}

\noindent Where $C_{1}$ is an undetermined constant. While the second term is somewhat mysterious, it is immediately apparent that the first, up to a constant, is the Gouy phase shift (\ref{gouy}) of the fundamental gaussian beam. Since $\hat{N}_{z}$ works on the whole extent of the beam, in retrospect, it is sensible that the phase take all possible values as the beam propagates from $-\infty$ to $\infty$. We do not claim that this is the whole story, it is likely the true conjugate is more complex. But nonetheless the results are compelling.\\

\noindent Some useful insights can be gained from this. The Gouy phase can be seen as resultant from the spatial confinement of the beam \cite{gp}. The Gouy phase as a ``focusing anomaly" then comes from the additional confinement of a beam as it moves through a focus. Thus, if we are to believe the above, the $n$ quantum number can be considered conjugate to the transverse spatial confinement of the photon. There is also experimental evidence for such an interpretation. Studies of the full spiral bandwidth of LG beams resulting from spontaneous parametric down conversion (SPDC) \cite{full1,full2,mario} show that there is a strong $n$-mode correlation between the signal and idler beams. The two daughter photons have the same $n$ quantum number with a high probability. This effect is the strongest when the SPDC crystal is very large when compared to the width of the pump beam. However, as the beam grows wider with respect to the transverse extent of the crystal the correlation becomes less and less distinct. In our terms, when the spatial confinement goes from being undefined (functionally infinite when the transverse extent of the crystal is much larger than the beam) to well defined (the case where the crystal's extent is on the same order as the beam width) the $n$ quantum number moves from being sharply defined to less defined. This is suggestive of an uncertainty relation between the two quantities \--- as exists between conjugate pairs.\\

\noindent There are some problems with this formulation, however. Firstly, there is the assumption that the commutation relation be of the given form. The complexity of other commutation relations between discreet values and continuous parameters (eg. OAM-azimuthal and number-phase) suggests that this is likely not generally the case. Secondly, in the paraxial regime the $z$-coordinate is regarded as a parameter, not an observable \--- in the same sense that time is a parameter of the Schr\"{o}dinger equation. \\

\noindent As stated earlier the results we present are not intended to be seen as a finished result, but instead to spark discussion and interest.\\

\noindent Continuing with the study of uncertainties as they pertain to the LG modes, it is interesting to examine what role $n$-modes play in the radial variance. The standard equation for the variance is given by $\triangle^{2}r=\langle r^{2}\rangle-\langle r\rangle^{2}$. Applying the well known orthogonality relations for Laguerre polynomials we find

\begin{eqnarray}
\triangle^{2}r=\frac{2n+l+1}{2}w_{z}^{2}.
\end{eqnarray}

\noindent We see that $n$, along with the OAM, quantizes the radial variance given a specific propagation distance (and thus beam waist).\\

\noindent There is another potentially interesting physical quantity the radial number has a direct effect upon \--- that is, the phase velocity of the LG photon. The phase velocity may be computed directly from the absolute transverse amplitude of a beam \cite{phase} using the equation

\begin{eqnarray}
c^{2}-v_{p}^{2}=\frac{\nabla^{2}_{t}E_{A}}{k_{o}^{2}E_{A}}v_{p}^{2}.
\end{eqnarray}

\noindent Where $v_{p}$ is the phase velocity, $c$ is the speed of light in vacuum, and $E_{A}$ is the phaseless electric amplitude of the field. Given the relationship between the transverse Laplacian and the $n$-operator we can show how $n$ and $l$ together quantize the phase velocity (at $z=0$ and $r=w_{o}$, for simplicity)

\begin{eqnarray}
c^{2}-v_{p}^{2}=-\frac{4v_{p}^{2}}{k_{o}^{2}w_{o}^{2}}(2n+l).
\end{eqnarray}

\noindent It is also potentially useful to point out a similarity between $\hat{N}$ and the Iwasawa decomposition in first-order optics (see, for example, Eq.(2.6) in Ref.\cite{IW}), which may have some deeper meaning.\\

\noindent We now return to the structure of the $n$-mode operator itself at $z=0$, Eq.(\ref{N0}), and its strong resemblance to the harmonic Hamiltonian. We can surmise that a system which itself is described by such a harmonic Hamiltonian will preserve the $n$ quantum number of a photon as it propagates through it. A graded index optical fiber (GRIN fiber) is one such system \cite{grin} (at least for the case of zero OAM). We conjecture then that LG photons in a GRIN fiber would preserve the orthogonality of all the $n$-modes for arbitrary propagation distances and offsets. This would be highly advantageous from the perspective of quantum multiplexing using the $n$ quantum number as well as other well-defined propagation invariant quantities of photons such as OAM and polarization for classical and quantum informatics tasks.\\

\section{Conclusions}

In this short, preliminary note we first ask the question ``What, physically, are the radial modes of the Laguerre-Gauss beams?" and then attempt to answer it. We develop a differential operator for the radial quantum number which has as its eigenstates the LG modes and the $n$-integer as its eigenvalues. We draw some connections between the $n$ quantum number and a few other physical quantities \--- most interestingly we show that the conjugate to $n$ may be related to the Gouy phase shift, which is related to the transverse spatial confinement potentially giving a novel, and potentially deep, interpretation of some experimental results of radial mode correlations from spontaneous parametric down conversion. We also show how a graded index optical fiber may intrinsically preserve the $n$ quantum number under propagation, which potentially opens up another high-dimensional Hilbert space for quantum communication and multiplexing.

\section*{Acknowledgements}

\noindent The authors would like to acknowledge Mario Krenn, Robert Fickler, Gabriel Molina-Terriza and Ebrahim Karimi for useful discussions and input. This work was supported by the ERC Advanced Grant QIT4QAD, and the Austrian Science Fund FWF within the SFB F40 (FoQuS) and W1210-2 (CoQuS).


\begin{thebibliography}{10}
\bibitem{A}L. Allen, M.W. Beijersbergen, R.J.C. Spreeuw, and J.P. Woerdman, Phys. Rev. A \textbf{45}, 8185 (1992).
\bibitem{rev}S. Franke-Arnold, L. Allen, and M. Padgett, Laser and Photonics Reviews \textbf{2}, 299 (2008).
\bibitem{karimi}E. Karimi, and E. Santamato, Opt. Lett. \textbf{37}, 2484 (2012).
\bibitem{gp}S. Feng, and H.G. Winful, Opt. Lett. \textbf{26}, 485 (2001).
\bibitem{full1}F.M. Miatto, A.M. Yao, and S.M. Barnett, Phys. Rev. A \textbf{83}, 033816 (2011).
\bibitem{full2}V.D. Salakhutdinov, E.R. Eliel, and W. L\"{o}ffler, Phys. Rev. Lett. \textbf{108}, 173604 (2012).
\bibitem{mario}M. Krenn, M. Huber, R. Fickler, R. Lapkiewicz, S. Ramelow, A. Zeilinger, arXiv:1306.0096v1 [quant-ph] (2013).
\bibitem{phase}Z. Chen, Y.K. Ho, P.X. Wang, Q. Kong, Y.J. Xie, W. Wang, and J.J. Xu, App. Phys. Lett. \textbf{88}, 121125 (2006).
\bibitem{IW}R. Simon, and N. Mukunda, J. Opt. Soc. Am. A \textbf{15}, 2146 (1998).
\bibitem{grin}G. Molina-Terriza, L. Torner, E.M. Wright, J.J. García-Ripoll and V.M. Pérez-Garc\'{i}a Opt. Lett. \textbf{26}, 1601 (2001).
\end{thebibliography}
\end{document}